# Where the Earth is flat and 9/11 is an inside job: A comparative algorithm audit of conspiratorial information in web search results

**Authors:** Aleksandra Urman[a]*, Mykola Makhortykh[b], Roberto Ulloa[c], Juhi Kulshrestha[d]

[a]*Department of Informatics, University of Zurich, Switzerland;* [b]*Institute of Communication and Media Studies, University of Bern, Switzerland;* [c]*GESIS - Leibniz-Institut für Sozialwissenschaften, Germany;* [d]*Department of Politics and Public Administration, University of Konstanz, Germany.*

*Corresponding author. Email: urman@ifi.uzh.ch*

**Abstract:** Web search engines are important online information intermediaries that are frequently used and highly trusted by the public despite multiple evidence of their outputs being subjected to inaccuracies and biases. One form of such inaccuracy, which so far received little scholarly attention, is the presence of conspiratorial information, namely pages promoting conspiracy theories. We address this gap by conducting a comparative algorithm audit to examine the distribution of conspiratorial information in search results across five search engines: Google, Bing, DuckDuckGo, Yahoo and Yandex. Using a virtual agent-based infrastructure, we systematically collect search outputs for six conspiracy theory-related queries ("flat earth", "new world order", "qanon", "9/11", "illuminati", "george soros") across three locations (two in the US and one in the UK) and two observation periods (March and May 2021). We find that all search engines except Google consistently displayed conspiracy-promoting results and returned links to conspiracy-dedicated websites in their top results, although the share of such content varied across queries. Most conspiracy-promoting results came from social media and conspiracy-dedicated websites while conspiracy-debunking information was shared by scientific websites and, to a lesser extent, legacy media. The fact that these observations are consistent across different locations and time periods highlight the possibility of some search engines systematically prioritizing conspiracy-promoting content and, thus, amplifying their distribution in the online environments.

**Introduction**

Web search engines (SEs) are crucial information gatekeepers in contemporary high-choice information environments (Van Aelst et al., 2017) with internet users turning to them on a daily basis (Urman and Makhortykh, 2021). At the same time, as demonstrated by a mounting body of evidence, search results can be inaccurate or biased (Kay et al., 2015; Kulshrestha et al., 2017; Makhortykh et al., 2020; Noble, 2018; Otterbacher et al., 2017). Still, search outputs are highly trusted by people and can influence their opinions on matters ranging from commercial brands to elections (e.g., Fisher et al., 2015; Nichols, 2017). Thus, malperformance of SEs can cause societal problems by leading, for example, to the spread of misinformation or of racial stereotypes (Noble, 2018).

While the explorations of bias in search results are increasingly common (see below), other forms of SE malperformance, in particular the one related to inaccurate search results, remain under-studied with a few notable exceptions (Bernstam et al., 2008; Bradshaw, 2019; Cooper and Feder, 2004). Unlike biased outputs, which tend to disproportionately amplify a particular point of view - e.g., by associating modern technology with Whiteness (Makhortykh et al., 2021a), - inaccurate outputs contain factually incorrect information (e.g., that the Earth is flat). Consequently, inaccurate outputs have higher potential for misinforming the users of SEs, which in some cases can pose a threat for their individual well-being as well as the society. It is particularly valid for outputs promoting conspiracy theories[1], which unlike other forms of incorrect or biased search outputs has so far received meager attention from the scholarly community. As shown by the ongoing COVID-19 crisis (European Commission, 2021), conspiracy theories diminish trust towards authorities and scientific community which can undermine societal cohesion and lead to radicalization, in particular at the time of crises.

In this paper, we address the above-mentioned gap by investigating the presence of content promoting conspiracy theories in web search results through a systematic comparative algorithm impact audit. We rely on virtual agent-based infrastructure to systematically collect search outputs for six conspiracy theory-related queries on five most popular SEs across three locations and two waves (in March and in May 2021). Out of six utilized queries, three correspond to specific conspiracy theories ("flat earth", "new world order", "qanon") - and are likely to be utilized by users interested in respective theories. Another three broadly refer to subjects about which many conspiracy theories circulate ("9/11", "illuminati", "george soros") - and can be utilized by users broadly interested in related topics, without specific interest in conspiracy theories. We then conduct a qualitative analysis of all retrieved results to establish their stance on conspiracy theories (e.g., promoting/debunking) and their sources (e.g., social media or scientific websites), and compare our observations across locations and time periods. With this paper, we contribute, first, to the body of research on the spread of conspiracy theories through online platforms by analyzing their presence in web search results which were not studied in this context before; and second, to the literature on algorithm auditing and quality of information provided by web search engines.

---

[1] We adhere to the formal definitions of "conspiracy theory" provided by the Merriam Webster dictionary: *"A theory that explains an event or set of circumstances as the result of a secret plot by usually powerful conspirators"; "a theory asserting that a secret of great importance is being kept from the public"* (Definition of CONSPIRACY THEORY, n.d.)

The rest of the paper is organized as follows: we first review the state of research on inaccurate and biased information in web search and on conspiracy theories online. Then, we build on this review to formulate concrete research questions and describe the methodology of our study in detail. Finally, we summarize our results and discuss their implications as well as the limitations of the current research.

**Web search results: biases, inaccuracies and their consequences**

In recent years a growing number of studies, primarily from the fields of computer science and communication science, have examined the presence of various biases in search results. To do so, researchers usually rely on a methodology called *algorithm impact auditing* - an investigation of the outputs of an algorithmic system - such as SE's information retrieval and ranking algorithms - with the aim to assess whether they contain biases or inaccuracies (Mittelstadt, 2016; see Bandy, 2021 for an overview of audits across different domains).

Algorithm audits in the context of web search have been largely focused on how specific biases lead to distortions in search results (Bandy, 2021). Scholars have found evidence of gender bias in image search (Kay et al., 2015; Otterbacher et al., 2017; Makhortykh et al., 2021a) and in Google's Knowledge Graph Carousel (Lipani et al., 2021), political bias in text search (Kravets and Toepfl, 2021; Kulshrestha et al., 2019), biases related to historical content in image search (Makhortykh et al., 2021b), and various forms of source-related biases (Diakopoulos et al., 2018; Haim et al., 2018; Kravets and Toepfl, 2021; Makhortykh et al., 2020; Puschmann, 2019; Trielli and Diakopoulos, 2019; Unkel and Haim, 2021; Urman et al., 2021a, 2021b). Further, several studies have documented information inequalities that can arise not only between users of different SEs but also between the users of the same SE due to search personalization or randomization (Hannak et al., 2013; Kliman-Silver et al., 2015; Paramita et al., 2021). Importantly, these inequalities can pertain to information crucial for individual and societal well-being, for instance concerning suicide helplines (Haim et al., 2017) or COVID-19 (Makhortykh et al., 2020).

While bias in web search received a lot of attention from scholars, there are only a few studies looking specifically at factually inaccurate information in search results (Bernstam et al., 2008; Bradshaw, 2019; Cooper and Feder, 2004). However, the possibility of web search promoting inaccurate information is concerning for a number of reasons. First, people highly trust the information they encounter through SEs, which are viewed as more reliable sources of information than legacy media (*2021 Edelman Trust Barometer | Edelman*, n.d.); consequently, the probability of individuals trusting inaccurate information acquired via SEs is high. Second, the mere act of using SEs increases individuals' confidence about their knowledge on the topic and motivates them to conflate external knowledge - that is the one available online - with their own (Ward, 2021). Third, people's search behaviour and interpretation of results are subject to selective exposure and confirmation bias that leads to them perceiving incorrect information that aligns with their existing attitudes as reliable and dismissing the results that do not align with their attitudes (Knobloch-Westerwick et al., 2015; Nichols, 2017; Suzuki and Yamamoto, 2020). Fourth, search results can sway people's opinions including on issues as important as voting preferences (Epstein and Robertson, 2015; Zweig, 2017) or commercial brand choices (Jansen et al., 2011). Fifth, inaccuracies in search results can proliferate to other socio-technical and informational systems that are built with reliance on SE outputs such as fake news detection algorithms relying on search outputs for the cues as to which information is false (Shim et al., 2021; Varshney and Vishwakarma, 2021).

Considering the potential harm which can be caused by search bias and factually inaccurate information, we suggest that comprehensive auditing of web search is of no less importance than the critical examinations of content appearing on legacy or social media. We strive to add to the body of literature on web search malperformance through an analysis of presence of conspiratorial information in search results. As we highlight below, this is a highly relevant subject due to the role the internet plays in the proliferation of conspiracy theories.

**Conspiracy theories in the online environments**

Multiple studies suggest that the rise of online platforms have been conducive to the broader circulation and proliferation of conspiracy theories (Stano, 2020). While this suggestion has been criticized (e.g., by arguing that conspiracy theories can convince only those who already have predispositions to believe in them, Uscinski et al., 2018), even critics agree that the internet amplifies the spread of conspiratorial information and facilitates exposure to it even for individuals who do not hold conspiratorial ideas (Uscinski et al., 2018).

While the spread of conspiratorial content by itself is unlikely to make the society as a whole to believe conspiracy theories, it is nevertheless concerning. It can increase belief in conspiracy theories in those parts of the population that have relevant predispositions. The spread of conspiratorial content might not lead to striking increases in the share of people who believe in conspiracy theories (Douglas et al., 2019), but it can have other problematic societal consequences such as decreasing prosocial behaviour and acceptance of science (van der Linden, 2015).

Broad circulation and "normalization" of conspiratorial information (Aupers, 2012) in the online environments has prompted increased scholarly interest in the phenomenon (for an overview see Douglas et al., 2019). Over the last decade, a range of studies has explored the way conspiracy theories are spread and discussed online (e.g., Bessi et al., 2015; Harambam, 2021; Lewandowsky et al., 2013; Mahl et al., 2021; Metaxas and Finn, n.d.; Mohammed, 2019; Röchert et al., 2022; Samory and Mitra, 2018; Uscinski and Parent, 2014; Wood and Douglas, 2013). Most of this research has focused on social media platforms which are deemed a favourable environment for the spread of conspiratorial content (Stano, 2020): not only do the (conspiracy theory-sharing) users cluster together there (Bakshy et al., 2015), but also false information tends to rapidly spread on social media (Vosoughi et al., 2018). Despite the importance of social media-focused research on how conspiracy theories are spread, we suggest that other online information retrieval and curation platforms - such as SEs - can contribute to the proliferation of conspiracy theories online. At the same time, to our knowledge, no systematic analysis of the presence of conspiratorial information in web search results has been conducted yet[2] and we aim to address this gap.

**Research questions**

---

[2] One recent study examined the proliferation of conspiracy theories related to COVID-19 in Google's autocomplete suggestions (Houli et al., 2021), but not search results.

Based on the research findings outlined in the previous section, we suggest that SEs can be conducive to the distribution of conspiratorial information and formation of conspiracy beliefs should such information appear in top search results. Unlike social media or online blogs, web search results have not been examined in the context of the distribution of conspiratorial information - despite numerous analyses showing that biased or not fully accurate results are not uncommon there. With this paper, we strive to partially fill the identified gaps in both - research focused on the presence of inaccurate information in web search and research on the spread of conspiracy theories online. Hence, our first research question concerns the general prevalence of conspiratorial information in top search results.

*RQ1: How prevalent is conspiratorial information in web search results?*

Because there are major discrepancies in the information provided by different SEs (e.g., Kravets and Toepfl, 2021; Makhortykh et al., 2020; Paramita et al., 2021), and in the results retrieved from different locations (Kliman-Silver et al., 2015), we find it worthwhile to address RQ1 from a comparative perspective to increase the robustness of the findings. Specifically, we compare the observations across different locations and on different search engines using the data collected at two different points in time. Further, we select two groups of conspiracy-related search queries, and expect to observe differences between the results retrieved for them, adding another comparative dimension to the analysis. Thus, we pose three sub-RQs in relation to RQ1:

*RQ1a: Does the prevalence of conspiratorial information in search results vary across different locations and time periods?*

*RQ1b: Does the prevalence of conspiratorial information in search results vary across SEs, and how?*

*RQ1c: Does the prevalence of conspiratorial information in search results vary across search queries, and how?*

Research, on one hand, demonstrates that SEs tend to over-represent certain source types while under-representing others (e.g., Puschmann, 2019; Haim et al., 2018), and, on another hand, suggests that conspiracy theories are mainly propagated on social media or small conspiracy-related websites (Douglas et al., 2019; Stano, 2020). Hence, we juxtapose the presence of conspiracies in web search with the prioritization of different sources and pose the following RQ:

*RQ2: What types of sources are prioritized by web SEs in relation to conspiracy theories?*

In the case of RQ2, we will also adopt a comparative approach similar to that of RQ1 - comparing the results for different engines, locations, waves and queries:

*RQ2a: Does the prevalence of sources of different types in search results vary across different locations and time periods?*

*RQ2b: Does the prevalence of sources of different types in search results vary across SEs, and how?*

*RQ2c: Does the prevalence of sources of different types in search results vary across search queries, and how?*

Additionally, we examine the relationship between the source type and the stance towards conspiracy theories. To do it, we formulate the last research question:

*RQ3: Are there differences in the share of conspiracy-promoting or conspiracy-debunking content coming from sources of different types?*

**Methodology**

*Data collection*

*Agent-based audit setup*. To collect the data we used agent-based algorithm impact auditing (see Ulloa et al., 2021). This type of auditing relies on automated agents, i.e., software that simulates user browsing behavior (e.g. scrolling web pages) and records the outputs. The benefit of this approach is that it allows controlling for search personalization (Hannak et al., 2013) and randomization (Makhortykh et al., 2020). Unlike human agents, automated agents can be easily synchronized (i.e., to isolate the effect of time at which the search actions are conducted) and deployed in a controlled environment (e.g., a network of virtual machines using the same IP range, the same type of operating system (OS) and the same browsing software) to limit the effects of personalization based on user-specific factors (e.g. location or OS type). Our approach is comparative in the sense that we assess the outputs of several SEs across multiple locations, time periods, and queries.

To increase the robustness of our observations, the data were collected over two collection rounds: March 18-19 and May 8-9 2021. From the technical standpoint, the data collection was organized as follows: each agent was made of two browser plugins (for either Chrome or Firefox desktop browsers). The first plugin (the *bot*) simulated human activity in the browser, in our case opening the search engine page, entering a query and scrolling down the first page of search outputs. The second plugin (the *tracker*) recorded html content appearing in the browser and sent it to the remote server.

The agents were deployed via Amazon Elastic Compute Cloud (EC2). We focused our exploration to the five biggest SEs by market share on desktop devices worldwide: Google, Bing, Yahoo, Yandex and DuckDuckGo (Desktop Search Engine Market Share Worldwide, n.d.). During the first round of data collection (March), we deployed 60 agents per location (see below), whereas for the second round we decreased this number to 30 agents per location because of budgetary limitations. The agents were equally distributed between the SEs (i.e., 12 agents per engine for the first round and 6 agents per engine for the second round). The majority of agents were able to perform their planned routine; the only major exception was Yandex, where multiple agents were blocked because of the aggressive bot detection mechanisms (e.g., frequent captchas) utilized by the search engine (hence, the absence of results for Yandex for some search queries).

*Search query selection*. We have collected search results for six queries that correspond to specific conspiracy theories ("flat earth", "new world order", "qanon") or to subjects related to conspiracy theories ("9/11", "illuminati", "george soros"). We specifically included these two groups of queries, because we suggest that they might be entered by users with two different intentions: those corresponding to the names of conspiracies are more likely to be entered by people already interested in a given theory, while the latter - by people merely interested in a certain topic. Thus if conspiratorial content is returned for the latter queries, it might lead to users' incidental exposure to conspiracies. Further, we expect to observe discrepancies between the two groups of queries with more conspiratorial content returned for the queries corresponding to specific theories and less for the subjects which are surrounded by conspiracy theories.

*Location selection*. We analyze the data collected for the five SEs, across three locations: two in the US (Ohio and California) and one in the UK (London). Our selection of locations is determined in part by the availability of different EC2 servers (see the description of technical implementation of the data collection below). Also, we utilized locations where English is the official language for the purposes of comparability - e.g., to make sure that any results we retrieve are due to the differences in the location, not to data voids in a specific language. Finally, we selected Ohio and California due to the differences in the ideological orientations of the two states (while California is a solid "blue" (Democratic) state in the US, Ohio is a "swing" state; we included it since there are no solid "red" states where EC2 servers are located); the inclusion of the UK server allows us to check whether our observations are country-specific. This comparative approach allows us to go beyond single-country observations that are common for algorithm audit studies (e.g., Haim et al., 2018; Puschmann, 2019; Urman et al., 2021a).

### Data analysis

After the data were collected, we extracted all unique URLs that appeared on the first page of search results for each engine and search query. We have focused on the first page of search results since people tend to perceive top search results as more credible and rarely look at/click on the results beyond the first page (Pan et al., 2007; Schultheiß et al., 2018; Unkel and Haas, 2017; Urman and Makhortykh, 2021). In total, there were 375 such unique URLs collected. The content under these URLs was then manually coded by two trained coders with the disagreements between the two resolved through consensus coding. The coding involved two variables, broadly corresponding to the RQs: conspiracy-related stance and type of source.

<u>The conspiracy-related stance variable</u> included the following categories:

- Debunks conspiracy: the content under the URL lists argument(s) why a conspiracy theory is not true
- No mention of conspiracy: there are no mentions of conspiracy theories in the source at all
- Promotes conspiracy: the content under the URL lists argument(s) why a conspiracy theory is true

- Mentions conspiracy: a conspiracy theory is mentioned, but there is no clear stance associated - e.g., no arguments for it being true or false OR arguments for it being true or false are presented in equal shares.

The type of source variable included the following categories:

- Reference website: e.g., online encyclopedias such as Wikipedia or Britannica
- Media: any media organization regardless of its ideological position or themes, except purely scientific news-focused sites (see the next category
- Science: scientific repositories such as JSTOR or news sites devoted exclusively to science such as ScienceNews
- Social media: e.g., Facebook pages, YouTube videos or Twitter accounts
- Conspiracy website: a website dedicated to the promotion of one specific conspiracy or of an array of conspiracy theories, such as a website of the Flat Earth Society
- Other: all other websites that do not fit into any of the previous categories

To assess the prevalence of conspiratorial content in search results (RQ1), we calculated the share of results for each stance towards conspiracy theories per condition (i.e., query, engine, location and collection round). To assess the prioritization of source types in relation to conspiracy content (RQ2), we first calculated the shares of results for each type per condition, and then the share of results with different stances towards conspiracy theories for each source type per location-round pair.

**Results**

*Prevalence of conspiratorial information in web search results*

We present the results of our analysis on the prevalence of information with different stances towards conspiracy theories (RQ1) per engine across all locations, collection rounds and queries in Figure 1, and disaggregated per query-engine-location in Figures 2-7 (Figures 2-4 correspond to the observations from March 2021; Figures 5-7 - from May 2021).

*Location-based and temporal differences*

The shares of content with different stances towards conspiracy theories vary widely across SEs and queries, but less so across different locations and waves. These observations indicate that our findings are not highly dependent on a specific location or time frame - at least when it comes to English-speaking locations and periods of observation that are relatively close to each other.

*Engine-based differences*

The only search engine that consistently did not return links promoting conspiracy theories is Google (see Figure 1), except for small shares of content promoting conspiracies in relation to the "new world order" query in May for the two US locations (Figure 5 and 6). The share of conspiracy-debunking content on Google was similar to that of other SEs for the most queries. At the same time, Google had the highest share of content that did not mention conspiracy theories at all, thus its users were less likely to be exposed to any conspiracy-related information.

The search engine with the highest proportion of conspiracy-promoting content was Yandex, while the remaining three engines - Bing, Yahoo and DuckDuckGo - were somewhere in-between Google and Yandex, with Yahoo having lower shares of content without mentions of conspiracy theories compared to the others but the highest proportion of conspiracy-debunking content. There were minor variations between the three with Yahoo having slightly lower shares of conspiratorial content. The similarities between Bing, Yahoo and DuckDuckGo are, perhaps, to be expected since Yahoo and DuckDuckGo are partially powered by Bing's search algorithms and results.

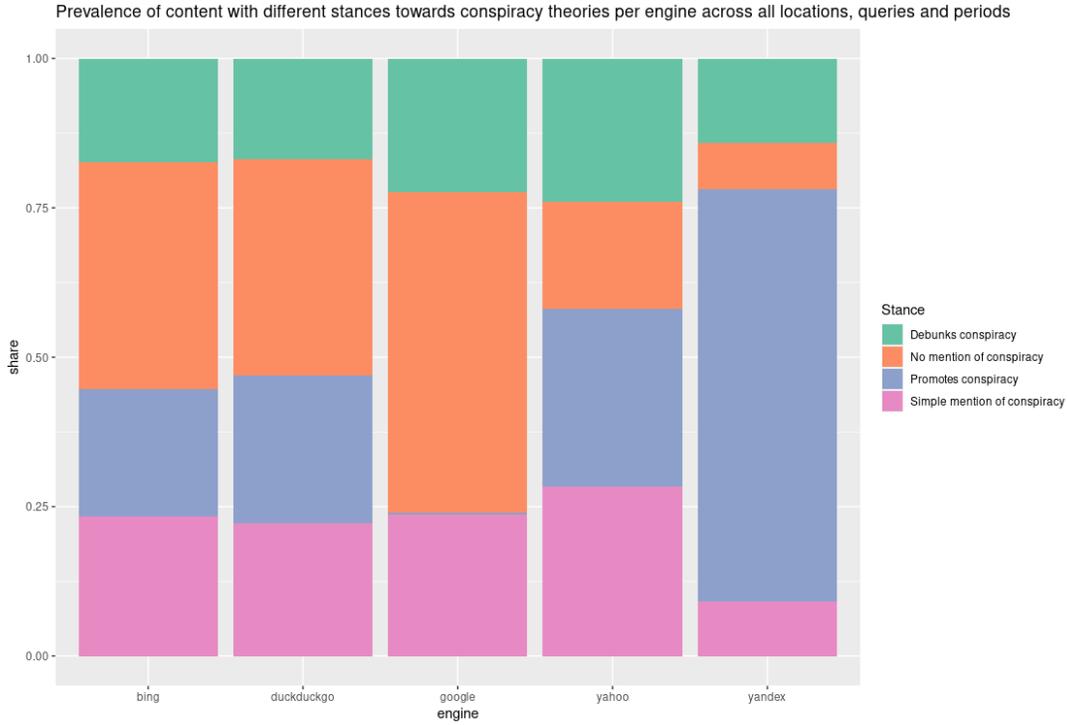

*Figure 1. Prevalence of results with different stances towards conspiracy theories per engine across all queries, locations and periods.*

*Query-based differences*

We observed evident discrepancies in the shares of content with different stances towards conspiracy theories across queries. Results for "9/11" contained the lowest shares of conspiracy-promoting content across all engines except Yahoo where, depending on the location and collection round, the share of conspiratorial content - e.g., that 9/11 was an "inside job" - reached close to 50%. No search engine returned results debunking conspiracy theories related to 9/11 but that arguably can be connected to the data void on this - i.e., there is content promoting either the official version or the conspiratorial one, with the former stating facts, but not debunking the latter.

The queries "george soros" and "new world order" are the other two queries for which relatively high shares of content that does not mention conspiracy theories were returned. For both queries, we observed comparatively low (<20% for "new world order" and <30% for "george soros") shares of content promoting conspiracy theories on all engines except Yandex. There is also a difference between the results returned for these two queries. For "george soros", more results debunking conspiracy theories were returned, with no results that would simply mention the theories. For "new world order", we observed few debunking results but relatively high proportions of results that simply mentioned the theory - i.e., that there is an emerging secret totalitarian world government - without a clear stance towards it.

Finally, the queries that attracted the highest shares of conspiracy-related content were "illuminati", "flat earth" and "qanon". For "qanon" we retrieved mostly content debunking the theory - with the exception of Yandex. Bing, DuckDuckGo and Yahoo contained comparatively low (up to 25%) proportions of conspiracy-promoting content. There was no content without any mentions of the conspiracy theory on either of the engines, which is perhaps to be expected given that "qanon" is a rather unambiguous term. With "illuminati" and "flat earth" the shares of conspiracy-promoting content were high on all SEs (except Google). Yandex, like with other queries, contained the highest shares of conspiracy-promoting links, whereas Bing, DuckDuckGo and Yahoo, depending on the location and the wave, displayed between 25% and 50% of conspiracy-promoting results. While with "flat earth" information debunking the conspiracy - i.e., listing arguments why the Earth is not flat, - was commonly displayed across all SEs, for "illuminati" conspiracy-debunking results were much less prevalent (except for Google).

Some of our expectations outlined in the "Methodology" section were not supported by the analysis. Notably, the search results regarding "illuminati" (a subject surrounded by many conspiracy theories) contained more conspiracy-promoting information than the results for "new world order" (query naming a particular conspiracy theory). Qualitatively, we have established that the non-conspiracy-related results displayed for "new world order" were referring to this concept in the context of international relations (e.g., Slaughter's (2012) article in Foreign Affairs that was featured in Google's search results). In the case of "illuminati", we attribute the high shares of conspiracy-promoting content to the wide spread of the related conspiracy theory and frequent references to it in popular culture (e.g., Dan Brown's books).

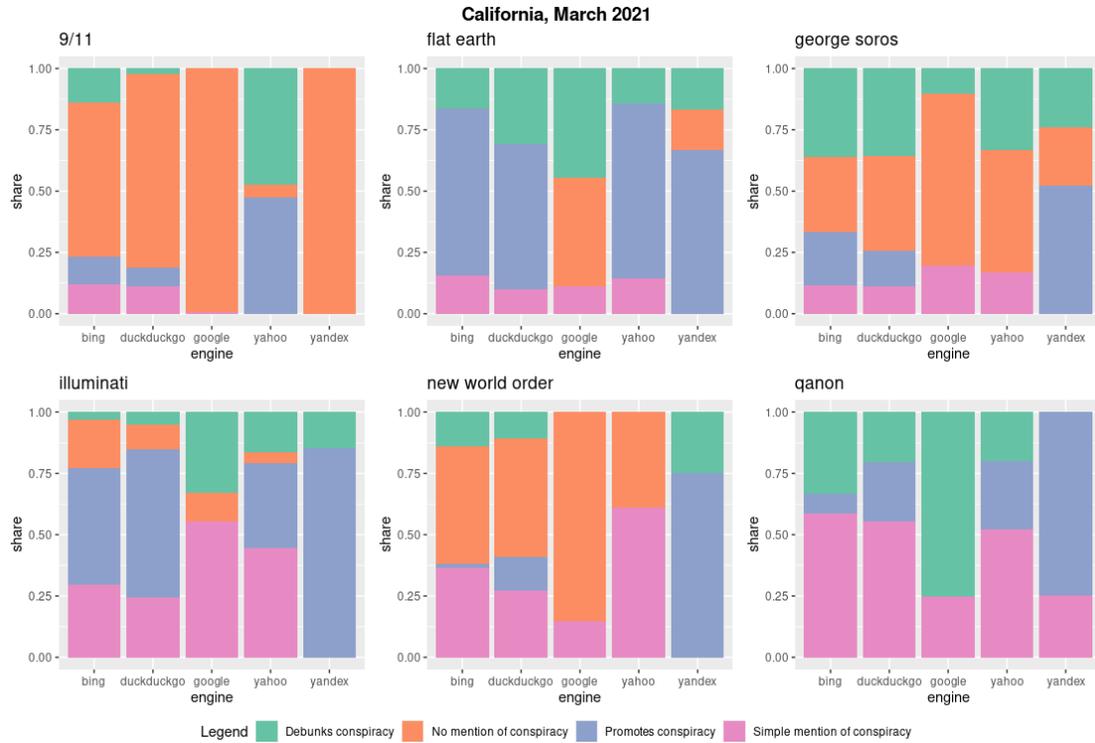

*Figure 2. Prevalence of content with different stances towards conspiracy theories per engine and query, California server, March 2021.*

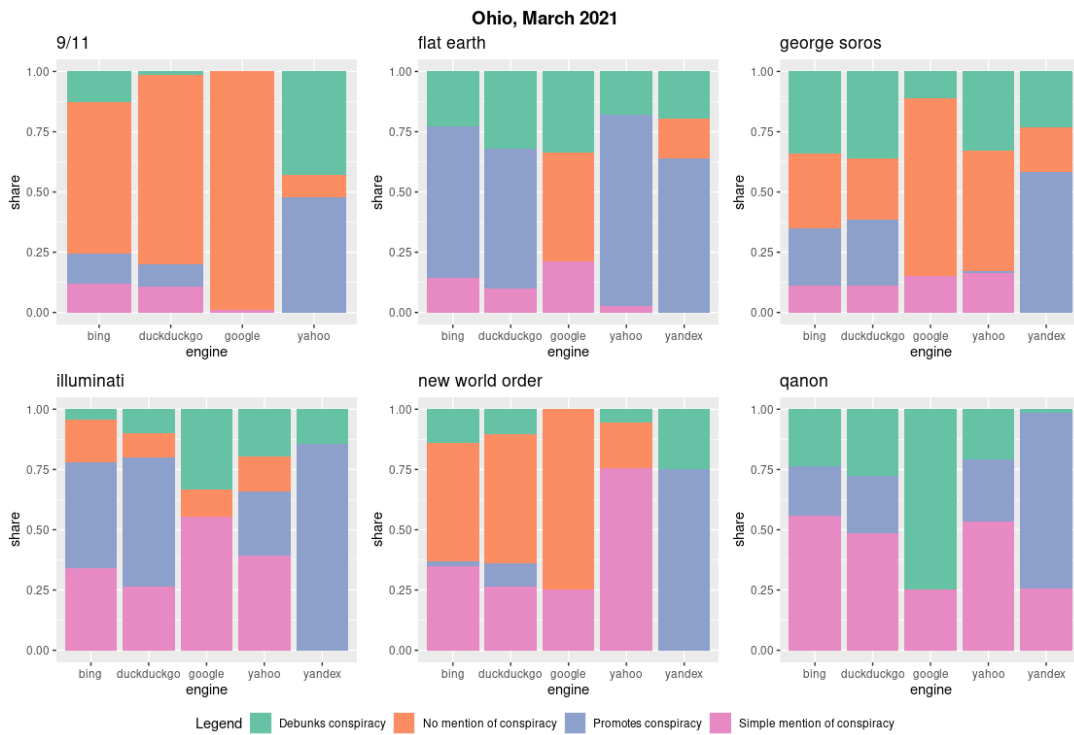

*Figure 3. Prevalence of content with different stances towards conspiracy theories per engine and query, Ohio server, March 2021.*

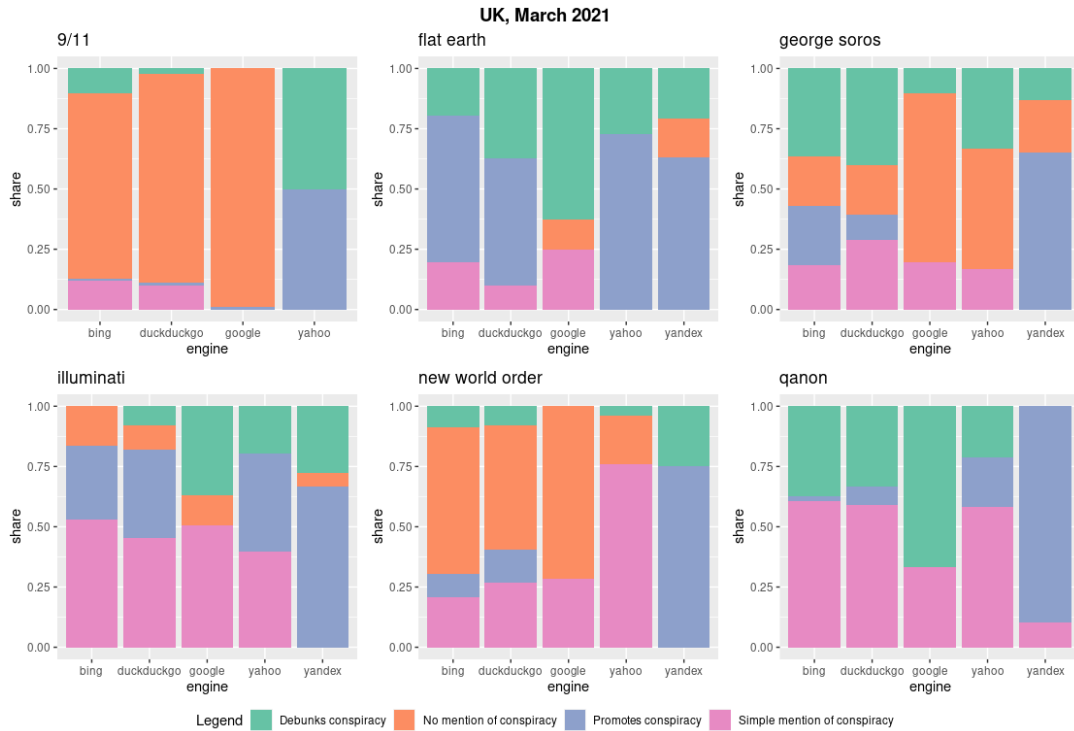

*Figure 4. Prevalence of content with different stances towards conspiracy theories per engine and query, UK server, March 2021.*

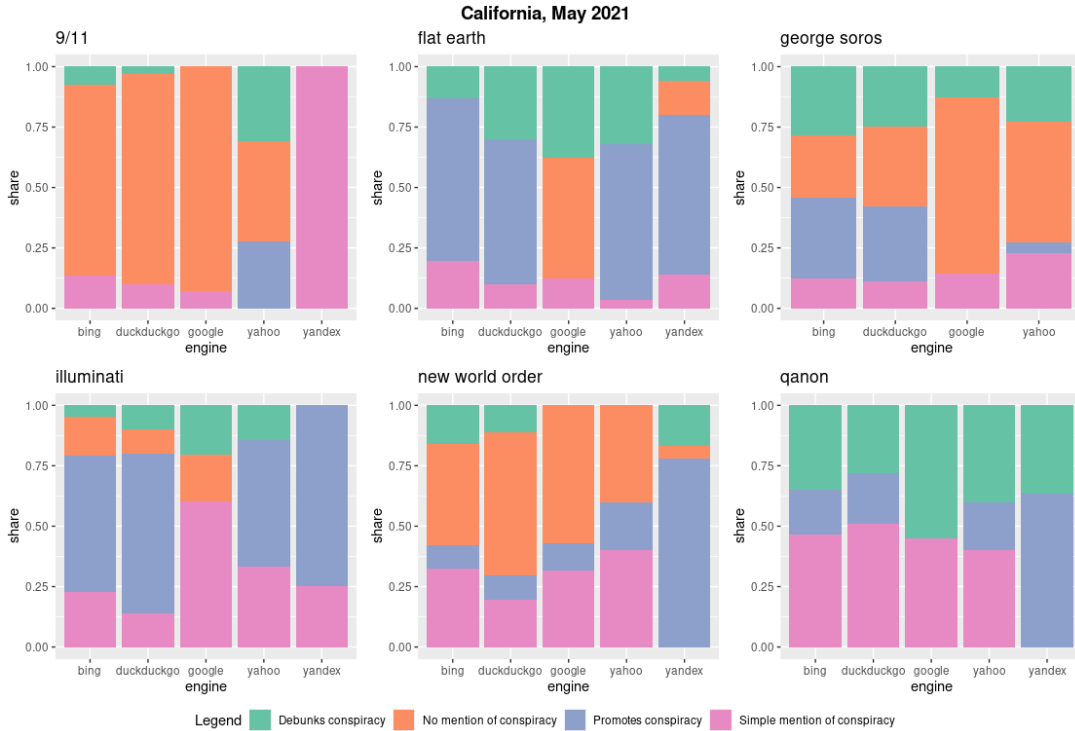

*Figure 5. Prevalence of content with different stances towards conspiracy theories per engine and query, California server, May 2021.*

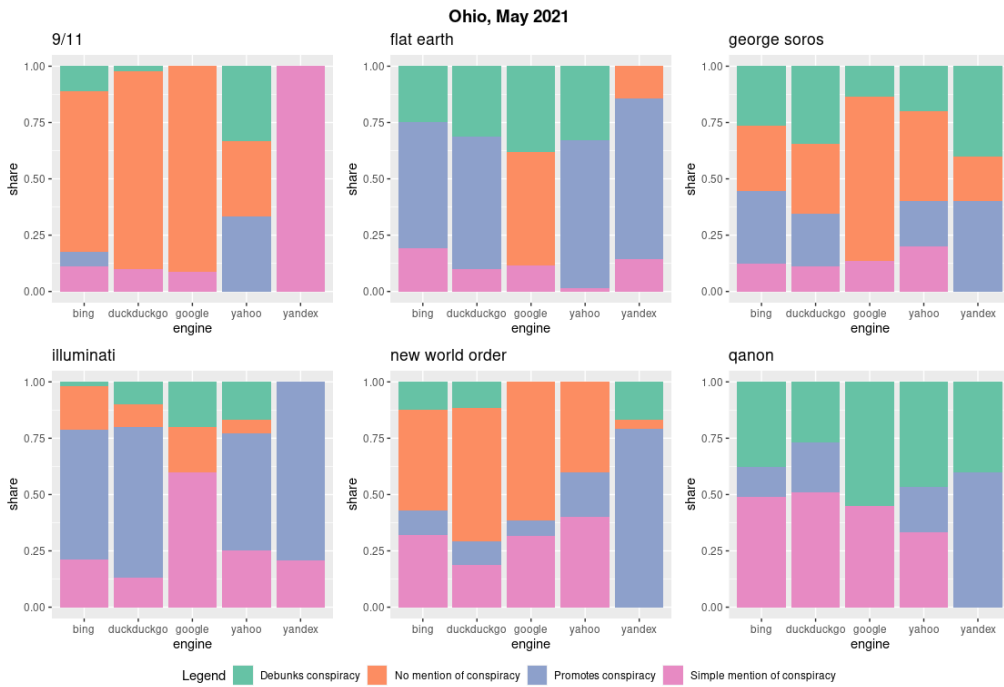

*Figure 6. Prevalence of content with different stances towards conspiracy theories per engine and query, Ohio server, May 2021.*

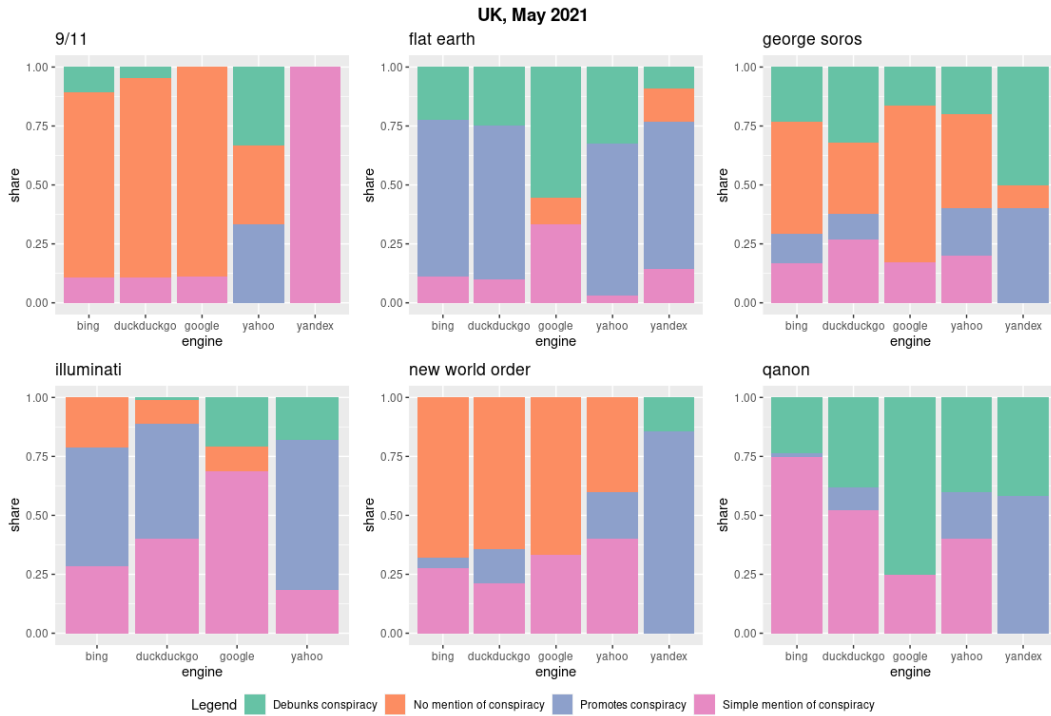

*Figure 7. Prevalence of content with different stances towards conspiracy theories per engine and query, UK server, May 2021.*

### *Prioritization of different types of sources in search rankings*

In relation to RQ2, we first present the results of the analysis of the shares of sources of different types across all queries-locations-waves (Figure 8), and then disaggregated per engine-query-location-wave combination (Figures 9-11 correspond to March 2021; Figures 12-14 correspond to May 2021). Finally, in Figure 15 we present the share of content with different stances per source type, aggregated across all SEs and queries for each of the two data collection rounds.

*Location-based and temporal differences*

Similar to the observations on the search results' stances towards conspiracy theories, we did not find major differences across locations and collection rounds. The only obvious difference relates to the prevalence of scientific sources: their share was higher in the UK than in the two other locations, and it was higher in March 2021 than in May 2021.

*Engine-based differences*

Our observations regarding the source types partially echo the findings on the content with different stances towards conspiracy theories reported above. Specifically, Google is the only search engine that did not return links to conspiracy-dedicated websites, while Yandex had the highest share of such links (Figure 8). Additionally, Google's results contained the biggest proportion of links to scientific sources, while those were absent on Yandex. In turn, Yandex was the engine with the highest share of links to social media in the results. Bing, Yahoo and DuckDuckGo contained similar shares of links to sources of different types with some differences for specific queries (e.g., Yahoo had fewer results from the media in response to "9/11" query than the other engines).

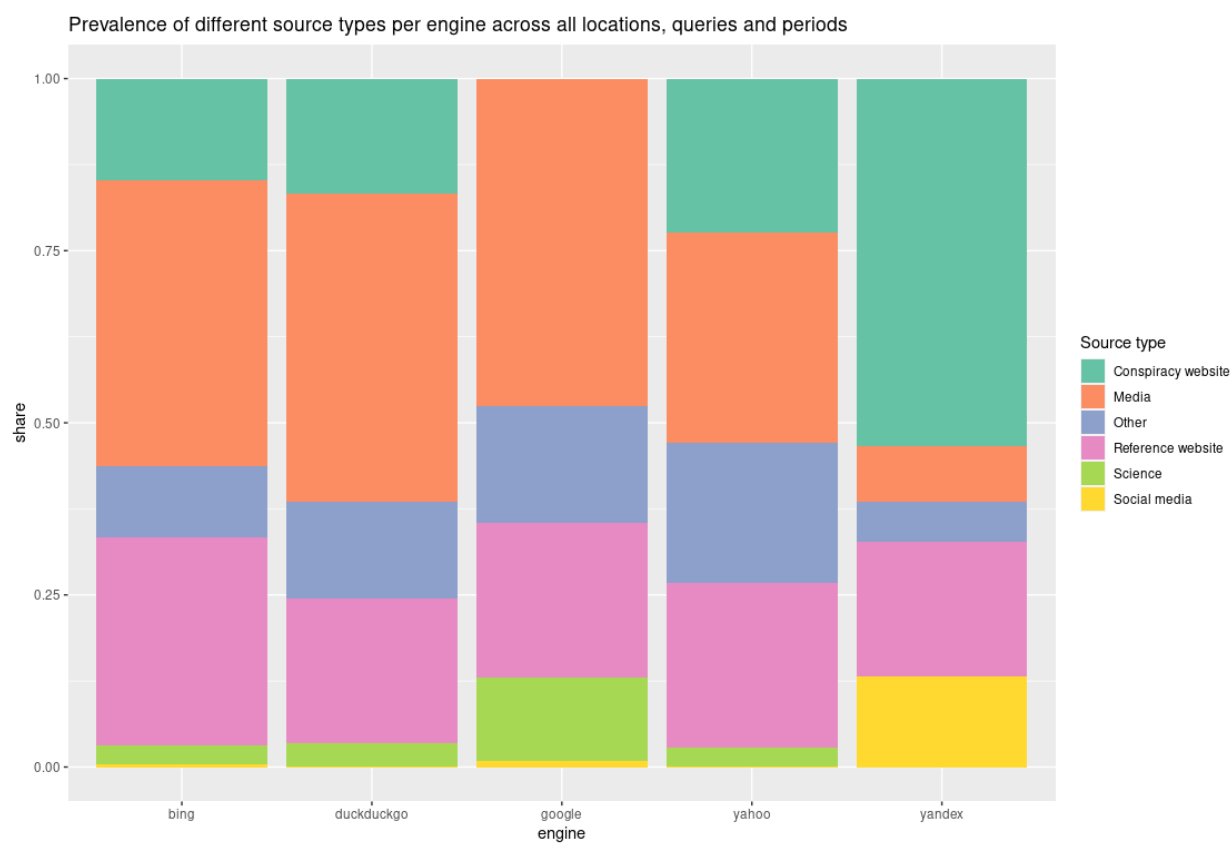

*Figure 8. Prevalence of different source types per engine across all queries, locations and periods.*

*Query-based differences*

For all queries except "flat earth" the outputs of SEs contained rather high shares of links to media and reference websites (e.g., Wikipedia pages), which is in line with what previous research found for queries related to COVID-19 (Makhortykh et al., 2020) or elections in the US (Kulshrestha et al., 2019; Urman et al., 2021). There were, however, some query-based discrepancies. For instance, for "qanon" the results had the highest shares of links to media, perhaps due to the intense media coverage of QAnon at the time of the data collection . The variations in the distribution of conspiracy websites across engines were similar to what we observe with regard to conspiracy stances: "flat earth"-related results had the highest shares of conspiracy websites, followed by "illuminati" and "qanon", while for the other three queries conspiracy websites were less prevalent.

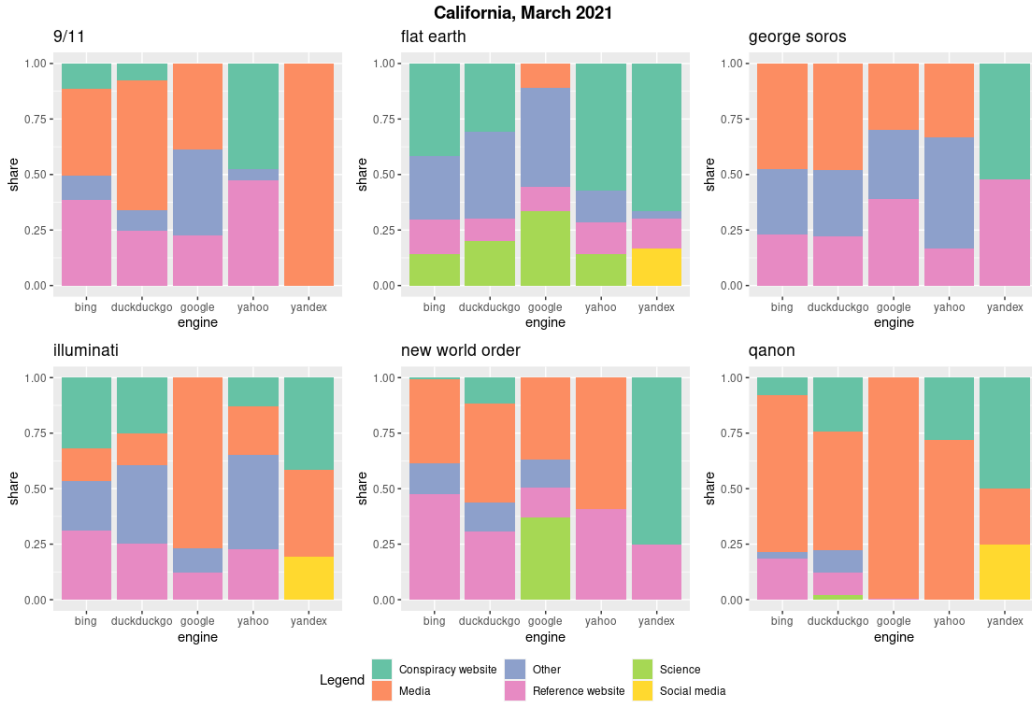

*Figure 9. Prevalence of different source types per engine and query, California server, March 2021.*

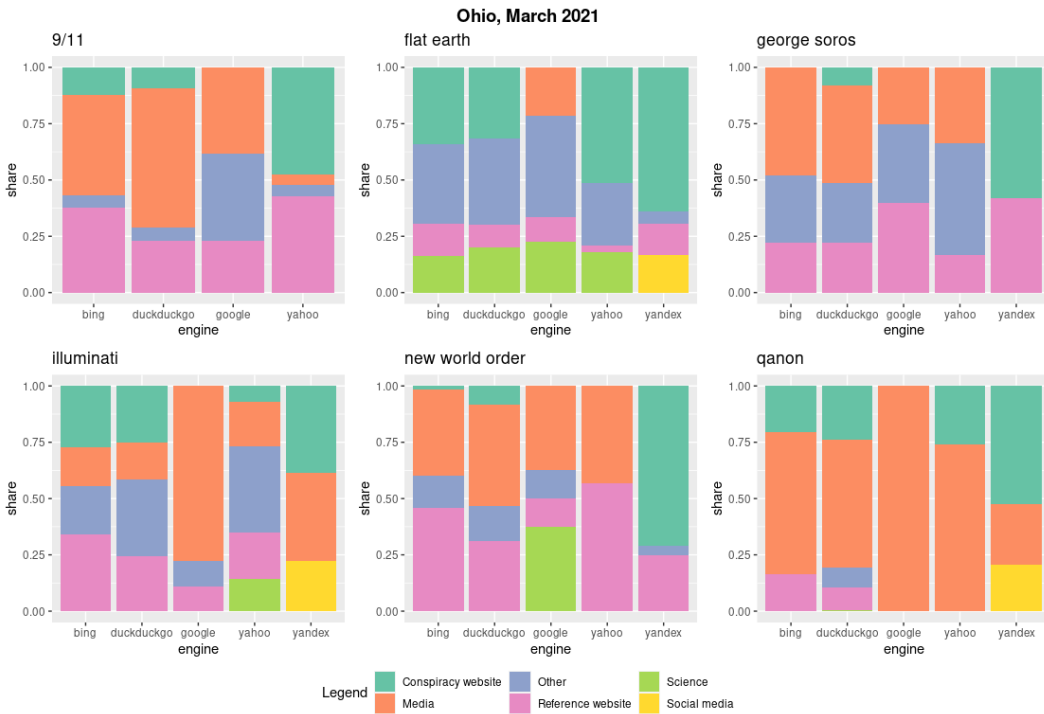

*Figure 10. Prevalence of different source types per engine and query, Ohio server, March 2021.*

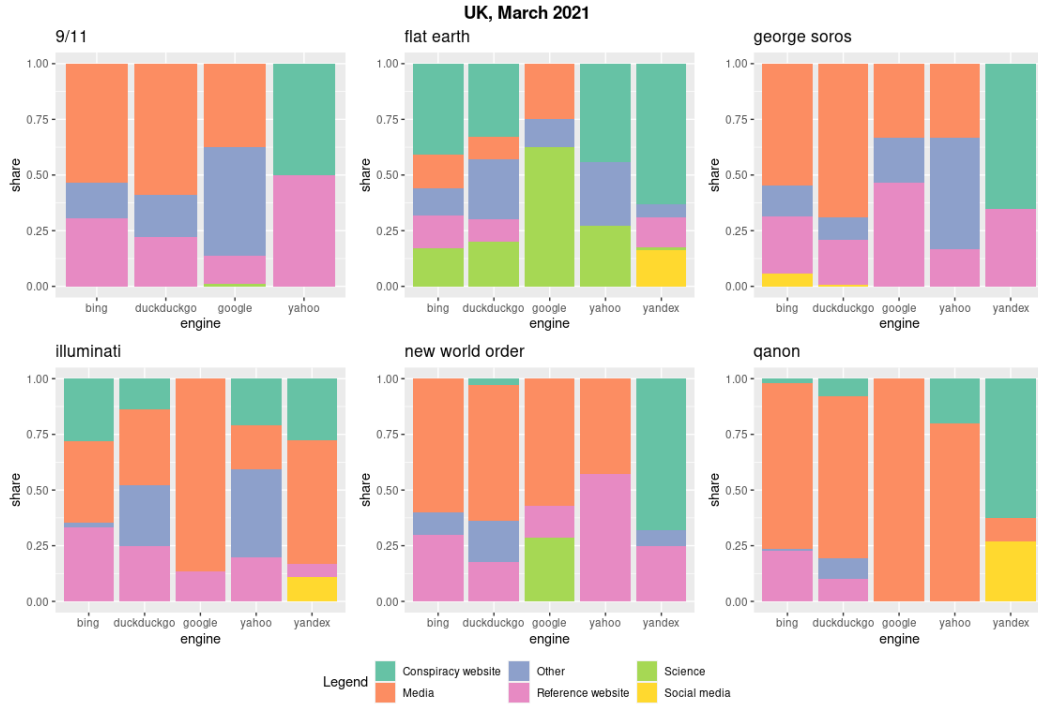

*Figure 11. Prevalence of different source types per engine and query, UK server, March 2021.*

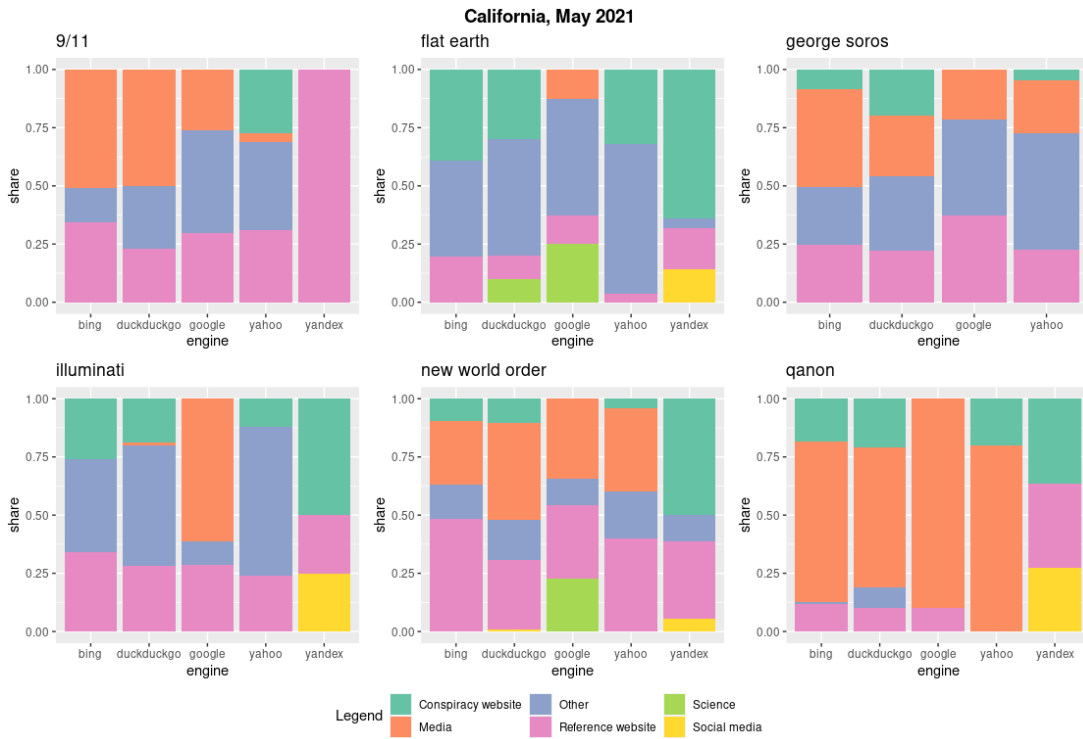

*Figure 12. Prevalence of different source types per engine and query, California server, May 2021.*

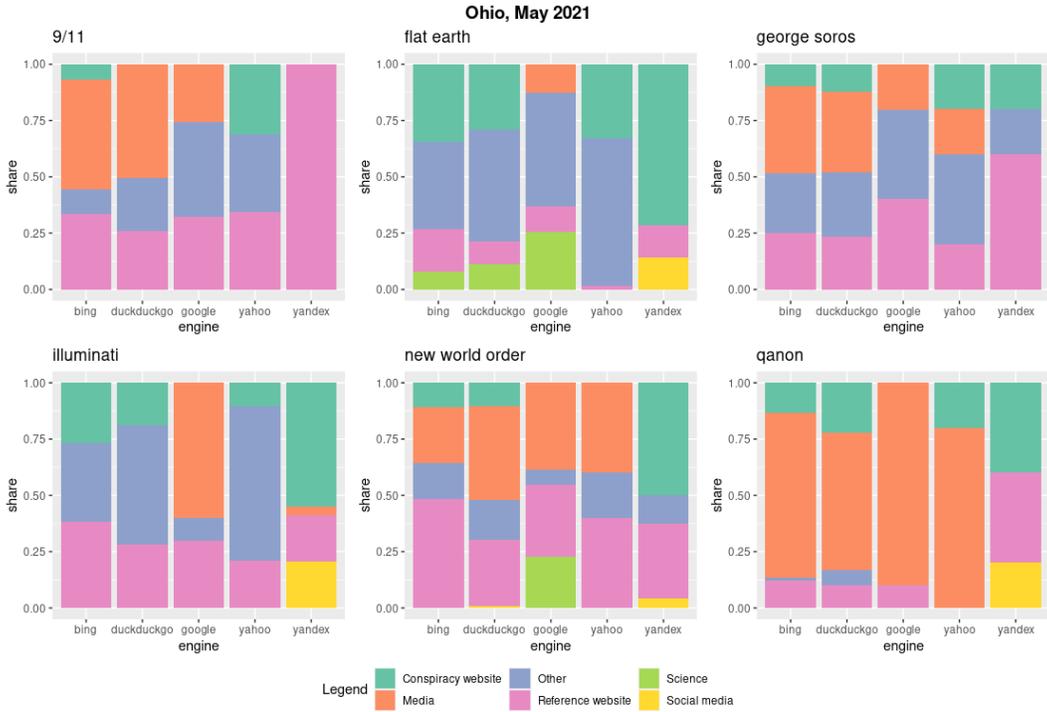

*Figure 13. Prevalence of different source types per engine and query, Ohio server, May 2021.*

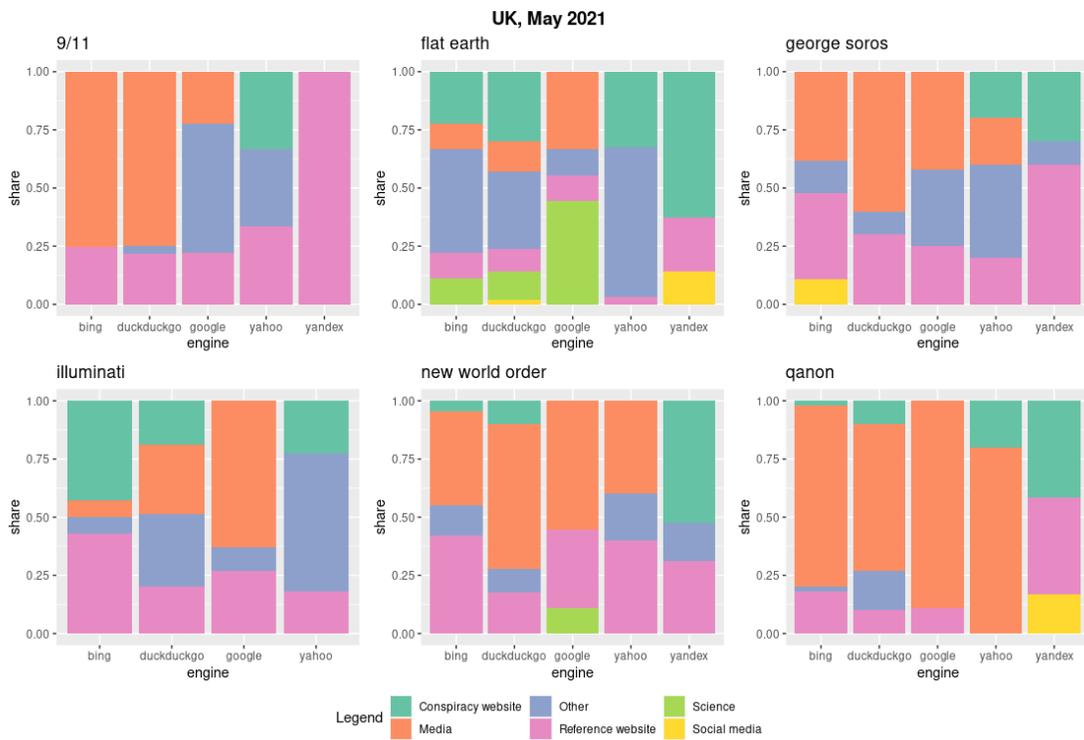

*Figure 14. Prevalence of different source types per engine and query, UK server, May 2021.*

*Source types vs stances towards conspiracy theories*

With regard to the shares of content with different stances towards conspiracy theories across source types (Figure 15), the observations within the two periods are rather similar, once again indicating the robustness of the findings against (short-term) temporal changes. Expectedly, all conspiracy-dedicated websites contained only information promoting conspiracy theories. Additionally, conspiratorial information was often present on social media that aligns with existing research on conspiratorial content spread online (e.g., Bessi et al., 2015; Stano, 2020). Among all other source types, the share of conspiracy-promoting content was the highest among websites in the "other" category. This included, for example, links to the webpages of individual books dedicated to specific conspiracy theories on Amazon webstore. There were very few media and reference sources that promoted conspiracy theories; instead, both source types predominantly either mentioned conspiracy theories or provided information unrelated to conspiracies. Media websites also contained high proportions of conspiracy-debunking content but the highest share of such information came from scientific websites.

The minor share of conspiracy theory-promoting information coming from scientific sources (see March 2021 observations in Figure 13) corresponds to one link to a peer-reviewed article from an academic journal in which the author postulated that International Relations scholars should question the official version regarding the September 11, 2001 attacks and backed this statement with the arguments coming from the "9/11 truth movement"[3]. For ethical reasons, we refrain from citing that particular paper here to not promote the conspiracy further.

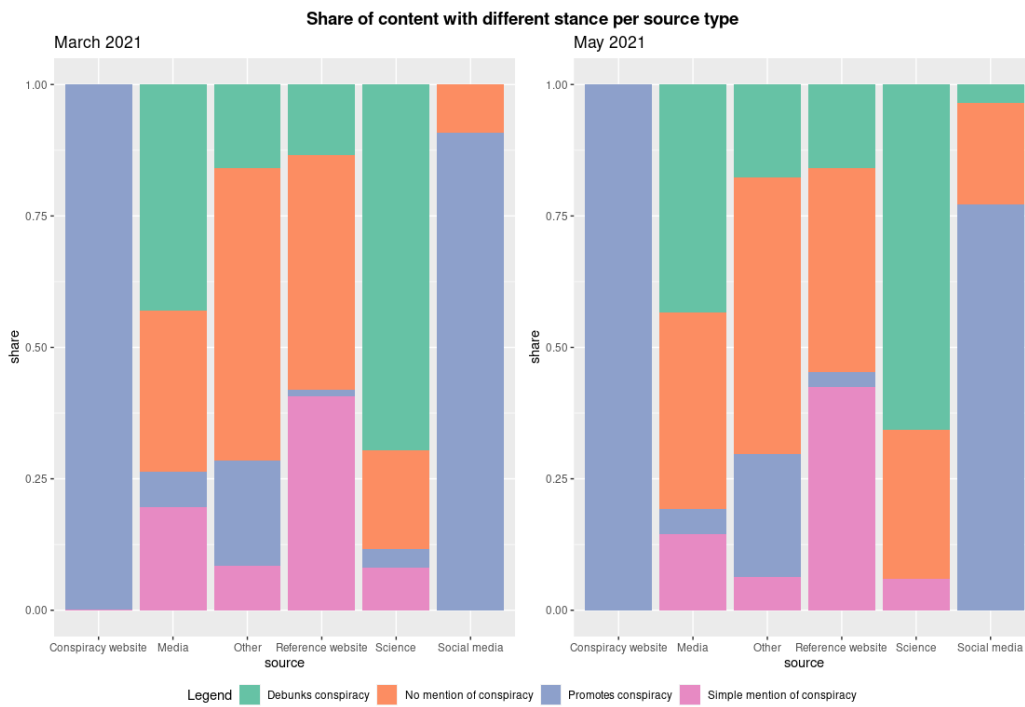

*Figure 15. Prevalence of content with different stances towards conspiracy theories per source type (aggregated across all search queries and engines for each wave).*

---

[3] "9/11 truth movement" refers to loosely connected individuals and groups that support conspiracy theories questioning the official version of the September 11, 2001 attacks (see more on the movement and related theories in Wood and Douglas, 2013)

**Discussion**

We have analyzed the presence of conspiratorial content in top (desktop) web search results across the five most popular SEs, three locations and two time periods for six conspiracy-related queries. Our findings contribute to both, the scholarship on inaccuracies and bias in web search, and research on the spread of conspiracy theories online. We also expect our observations to be robust considering there were only minor differences between locations and observation rounds.

Our analysis demonstrates that while there are large query-based differences, conspiracy-promoting content routinely resurfaces in the first page of results on all SEs except Google. Furthermore, Google's results also contain the highest proportion of scientific sources debunking the conspiracy theories. We suggest that the absence of conspiratorial information in Google's results is attributed to the company actively trying to reduce the share of misinformation and bias in its search outputs (Kayser-Bril, 2020). The company even explicitly covers the issue of conspiratorial content (including a specific example related to the 9/11 conspiracies) in its guidelines for search results' quality raters (Google, 2021).

Another finding concerns the types of web resources prioritized by different search engines. Besides Google, all other search engines we examined link to conspiracy theory-dedicated websites with Yandex being the engine with the highest share of links to conspiratorial content. At the same time, most content on all engines (except Yandex) comes from media and reference websites which tend to contain little conspiracy-promoting content. In line with the previous findings (e.g., Douglas et al., 2019), we observe that conspiratorial content is found mainly on dedicated niche websites or social media, but is rare on broad-reaching sites such as media, scientific and reference websites. The latter observation highlights that one of the ways to cull dissemination of conspiratorial information on search engines can be to implement domain-based filters de-prioritizing resources known to be involved in promoting such information, even though the long-term resilience of this solution can be questioned.

Filtering out conspiracy-promoting websites from the top search outputs is important for (at least) two reasons. While it can be assumed that only users who are already interested in conspiratorial content purposefully navigate to the dedicated niche websites (Douglas et al., 2019), their appearance in top search results, especially for queries that do not denote conspiracy theories per se (e.g., "9/11" or "george soros") can potentially lead to incidental exposure to conspiracy theories. This is troubling: given people's high trust in search results (*2021 Edelman Trust Barometer | Edelman*, n.d.), conspiratorial information found by users merely exploring the topic in top search results can induce the formation of conspiratorial beliefs. This might be especially the case if users have limited knowledge on a given topic - which is partially implied by them turning to SEs to explore it - as lower knowledge on a subject is associated with higher beliefs in related conspiracy theories (Sallam et al., 2020).

In the cases when the exposure to conspiratorial content via search engines is not incidental - i.e., if a person already interested in a conspiracy theory searches for related content online (e.g., "flat earth" or "illuminati")  - the appearance of conspiracy-promoting information in top search results is also concerning. If such a person is simply interested - but not yet believing - in conspiracy theories, conspiratorial content coming from a highly trusted source can foster conspiratorial belief development. If that person already, at least partially, believes in a conspiracy theory, their belief can be reinforced by conspiracy-promoting web search results due to high public trust in SE results coupled with confirmation bias (Knobloch-Westerwick et al., 2015; Nichols, 2017; Suzuki and Yamamoto, 2020). Due to the latter, the presence of even a single conspiracy-promoting result in top results might be enough to reinforce conspiracy beliefs.

**Limitations**

One limitation of the present study is that our selection of queries does not cover the whole range of existing conspiracy theories (Douglas et al., 2019) or all possible queries regarding a specific conspiracy theory, however, encompassing all theories or queries is arguably impossible, and our study should be treated as a first step towards more comprehensive analyses of conspiratorial information spread via SEs. Though the number of included theories is limited, we suggest that our approach is already more comprehensive than that of the majority of studies that focus on the spread of one specific theory online (with notable exceptions such as Mahl et al., 2021). As there is no established resource regarding the popularity of certain theories worldwide, our selection of queries was based on the internal discussions between authors, during which we tried to select the theories that are, first, in broad circulation and, second, that are "popular" worldwide, not within a single country.

Another limitation of the present study is our focus on just three anglophone Western locations and on one language - English. This makes our analysis Western- and English-centric. We aim to address this in future work by conducting studies on the topic with broader linguistic and geographic focus.

Finally, while we collected the data over two collection rounds to increase the robustness of the findings, these rounds were relatively close to each other. In the future, research infrastructure that enables more longitudinal data collection will be established to trace long-term changes (or their lack) in SEs' outputs.

**Conclusion**

We found that most of the search engines display conspiracy-promoting results, though the share of such results varies across specific conspiracy-related queries. In our sample most conspiracy-promoting results came from social media platforms and dedicated conspiracy websites, while debunking information was found predominantly on scientific websites and, to a smaller extent, on legacy media. Our observations are robust across several locations and two time periods. The good news is that Google - the search engine with the biggest market share - has managed to mitigate the problem to a few isolated instances. We suggest that the example of Google shows that conspiratorial results can be effectively handled to become less prevalent in top SE outputs, and that other SEs should follow suit and put the results they provide under higher scrutiny not only with regard to conspiracy theories but other types of inaccurate and/or biased information. This is especially relevant and timely as of now, when radical groups are attempting to create an alternative tech ecosystem and, among other, migrate from Google to DuckDuckGo, accusing the former of censorship and reinforcing the prioritization of far-right and/or conspiratorial sources on the latter (*Diggit Magazine*, 2020). It shows the potential for ideologically charged hijacking of smaller SEs that can influence their outputs. Against this backdrop, it is crucial to assure the quality of information SEs provide to users through both SEs' own internal monitoring and external audits such as the one conducted in the present study.